# Performance of Angle of Arrival Detection Using MUSIC Algorithm in Inter-Satellite Link

Cahya Budi Muhammad[1], Heroe Wijanto[2], Antonius Darma Setiawan[3]

[1,2,3] School of Electrical Engineering - Telkom University, Bandung, Indonesia

[1]cahyabudi@student.telkomuniversity.ac.id, [2]heroe@telkomuniversity.ac.id, [3]adsetiawan1701@gmail.com

*Abstract*— An attitude of satellite is not always static, sometimes it moves randomly and the antenna pointing of satellite is harder to achieve line of sight communication to other satellite when it is outage by tumbling effect. In order to determine an appropriate direction of satellite antenna in inter-satellite link, this paper analyze estimation performance of the direction of arrival (DoA) using MUSIC algorithm from connected satellite signal source. It differs from optical measurement, magnetic field measurement, inertial measurement, and global positioning system (GPS) attitude determination. The proposed method is characterized by taking signal source from connected satellites, after that the main satellite processed the information to obtain connected satellites antenna direction. The simulation runs only on the direction of azimuth. The simulation result shows that MUSIC algorithm processing time is faster than satellite movement time in orbit on altitude of 830 km with the period of 101 minutes. With the use of a 50 elements array antenna in spacing of 0.5 wavelength, the total of 20 angle of arrival (AoA) can be detected in 0.98 seconds of processing time when using MUSIC algorithm.

## I. INTRODUCTION

Attitude of communication satellites, especially direction of antenna pointing of satellite, must point toward to their satellites to obtain a better data link. The attitude of communication of satellite can be determined by measurements from attitude sensors. Attitude sensors produce the projection of the reference vector in the sensitive direction of attitude [1][2][3]. The used vectors are defined by stars, the sun, the earth, geomagnetic field, inertial space, and GPS satellite [3][4][5]. Most of communication satellites achieve attitude determination by corporately using different attitude references to prevent a system failure. For example, a three-axis gyroscope, a three-axis magnetometer, and a coarse horizon sensor are used for attitude references of iridium satellites [1][2]. The attitude of Globalstar satellites is determined by four Sun sensors, a horizon sensor, a three-axis magnetometer, and GPS [2].

Previous work about attitude determination with DOA estimation have been done by using ground signal source (a ground station or a ground mobile user) as a reference signal source and as the observation of the Extended Kalman Filter (EKF) for attitude determination of a communication satellite, where DOA attitude measurement equation is derived on the basis of the ground signal source location, the satellite orbit position, and the space geometric information between the signal source and the smart antennas. The results show that estimation accuracy increased in term of increasing SNR and number of snapshots, where the SNR simulated between -14 dB to 14 dB, and it give the best results for 14 dB. The simulation is analyzed in pitch, yaw, roll angle [8]. The second one is about direction finding to be used in a drone surveillance system using a sparse de-noising auto-encoder (SDAE) based deep neural network. The system comprises a single channel receiver and a directional antenna array. The receiver sequentially activates each antenna in the array and measures the received power level. The power measurements corresponding to each switching cycle are fed to the proposed deep network. Then, it performs direction finding by exploiting the sparsity property of the incoming drone signal, and the gain variation property of the directional antenna array. It has been proven that phase synchronization mechanism, antenna gain calibration mechanism, or profound knowledge about the antenna radiation pattern are not effective for this single channel implementation [11].

The third and the last one is finding direction of arrival estimation (DoA) using sparse reconstruction gives advantage on minimizing the number of required samples. Among available sparse reconstruction schemes, angle sparsity has shown a favorable advantage as it requires fewer samples compared to other schemes. Previous researches on angle sparsity utilized an exhaustive scanning on every possible arrival angles. This technique leads to a problem of large sensing matrix A. The result proves that partial scanning (i.e. non-exhaustive search) also gives similar accurate result. The advantage of this scheme is smaller sensing matrix. In simulation, this scheme requires sensing matrix six times less than the exhaustive search with similar accuracy, and it is potential for practical application of DoA estimation based on sparse reconstruction [12].

The work in this paper is based on a developing low earth orbit (LEO) satellite, where the information exchange between one main satellite to other connected satellites, and the information of angle of arrival can be obtained by array signal processing. The direction of arrival (DOA) estimation uses the data received by the smart array to estimate the direction of the signal source. The DOA contains information about angle of arrival that can be related to direction of pointing antenna from each connected satellite using MUSIC algorithm, where the algorithm is efficient enough because it is simple and based on eigen structure method [9]. A DOA estimation of antenna pointing satellite measurement equation is derived on the basis of the connected satellites orbit position. the performance of MUSIC algorithm is analyzed in detail. Then, a pointing antenna satellite determination is presented using azimuth direction of connected satellite. Finally the proposed method of antenna pointing of connected satellite determintaion is analyzed in term of Peak MUSIC (PMUSIC) and accuracy by exploring influences of maximum angular beamwidth, the number of antenna, the length of spacing between antenna, variation of frequency operation in inter-satellite links, the number of angle of arrival that can be related to number of connected satellite, and input signal-to-noise ratio (SNR) for each connected satellites.

## II. SIGNAL MODEL

### A. Signal Analysis

Consider Fig. 1, a narrowband signal $\bar{s}(t)$ with operating frequency $\omega_0$ arrives from angle $\theta$ and $\phi$ with $z$-axis and $x$-axis plane respectively. The narrowband signal can be expressed as

$$\bar{s}(t) = u(t)\cos(\omega_0(t) + v(t)), \quad (1)$$

where $u(t)$ and $s(t)$ is a variation of function from time which define as an amplitudo and phase. A delay depends on relative position from each sensor and against angle of arrival (AOA). If taking cartesian coordinate system as a reference location, and antenna element $i$ which related to signal on reference location can be expressed as [6]

$$\tau_i = -\frac{(x_i \sin\theta\cos\phi + y_i \sin\theta\sin\phi + z_i \cos\theta)}{c}, \quad (2)$$

where $c$ is a light velocity. Because the signal is a complex narrowband signal, the effect of delay propagation $\tau_i$ makes a phase shift $\xi_i = -\omega_0 \tau_i$ can be represented by

$$s(t - \tau_i) = s(t)\exp(j\xi_i) = s(t)\exp(-j\omega_0 \tau_i), \quad (3)$$

where phase shift $\xi_i$ given by [5]

$$\begin{aligned}\xi_i &= \frac{\omega}{c}(x_i \sin\theta\cos\phi + y_i \sin\theta\sin\phi + z_i \cos\theta)\\ &= \frac{2\pi}{\lambda}(x_i \sin\theta\cos\phi + y_i \sin\theta\sin\phi + z_i \cos\theta)\end{aligned} \quad (4)$$

The signal received by array can be expressed in a vector [9]

$$x(t) = \begin{bmatrix} x_1(t) \\ x_2(t) \\ \dots \\ x_M(t) \end{bmatrix} = \begin{bmatrix} e^{j\xi_1} \\ e^{j\xi_2} \\ \dots \\ e^{j\xi_M} \end{bmatrix} s(t). \quad (5)$$

Vector from $x(t)$ always referenced as input vector array data or illumination vector. In (5), phase shift separated from signal $s(t)$ because of existance of spatial separation between array antenna. For general cases, array element will have a direction and frequency response which depends on each element. That mattter can be modelled with applying different gain and phase for vektror from each elemen in (5). If the direction and frequency dependant gain and phase from antenna elemen $i$ denoted as $g_i(\omega, \theta, \phi)$, the signal in output array can be expressed as [5]

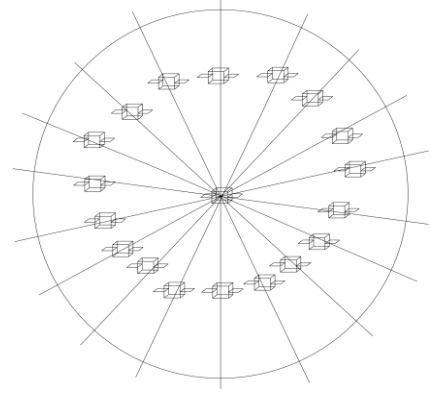

Fig. 1. Inter-Satellite-Link Model

$$x(t) = \begin{bmatrix} x_1(t) \\ x_2(t) \\ \dots \\ x_M(t) \end{bmatrix} = \begin{bmatrix} g_1(\omega,\theta,\phi)e^{j\xi_1} \\ g_2(\omega,\theta,\phi)e^{j\xi_2} \\ \dots \\ g_M(\omega,\theta,\phi)e^{j\xi_M} \end{bmatrix} \quad (6)$$

$$= \alpha(\omega,\theta,\phi)s(t)$$

where the vector

$$x(t) = \begin{bmatrix} g_1(\omega,\theta,\phi)e^{j\xi_1} \\ g_2(\omega,\theta,\phi)e^{j\xi_2} \\ \dots \\ g_M(\omega,\theta,\phi)e^{j\xi_M} \end{bmatrix} \quad (7)$$

$$= \begin{bmatrix} g_1(\omega,\theta,\phi)e^{j\frac{\omega}{c}(x_1\sin\theta\cos\phi + y_1\sin\theta\sin\phi + z_1\cos\theta)} \\ g_2(\omega,\theta,\phi)e^{j\frac{\omega}{c}(x_2\sin\theta\cos\phi + y_2\sin\theta\sin\phi + z_2\cos\theta)} \\ \dots \\ g_M(\omega,\theta,\phi)e^{j\frac{\omega}{c}(x_M\sin\theta\cos\phi + y_M\sin\theta\sin\phi + z_M\cos\theta)} \end{bmatrix}$$

is called steering vector. Equation (7) represent general form from steering vector from an array. Steering vector is a response function from each element, array geometry, frequency signal and angle of arrival.

### B. MUSIC Method

Multiple Signal Classification (MUSIC) is a technique that quite popular which used in direction of arrival estimation. It can be concluded that DOA estimation as a job to estimate angle of arrival which the location is unknown to antena receiver with some signal processing technique.

MUSIC method is relative simple and eigenstructure method which is efficient from DOA estimation. The technique can be done by doing eigenvalue decomposition from matrix correlation estimation from array or singular decomposition value, which column N represent N snapshots from array vector signal [9].

Matrix X from array sensor is l with n matrix where lth is total from antenna sensor and n represent total snapshots which taken. The matrix can be formulated as [10]

$$X^T = \begin{bmatrix} x_1 & \dots & x_l \end{bmatrix}. \quad (8)$$

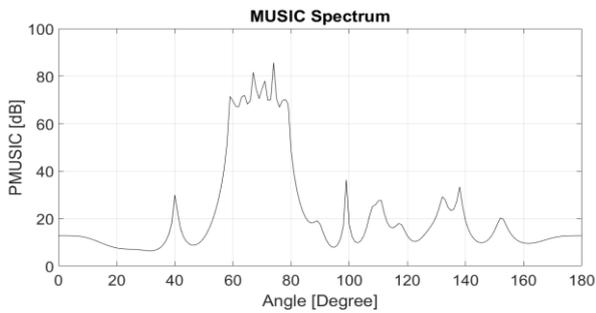

(a)

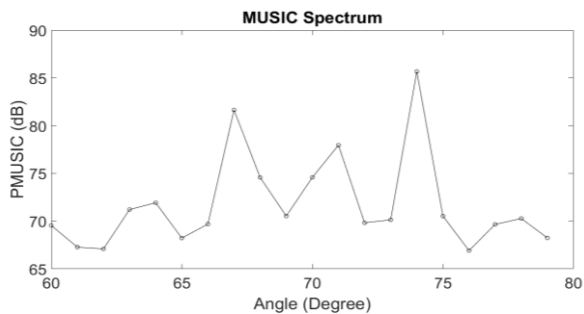

(b)

Fig. 2. Maximum Angular Beamwidth Simualtion Results at: (a) 0 – 180 Degree. (b) 60 – 80 Degree

The total signal that induced in element l from receiver array can be formulated as [7]

$$x_l = \sum_{k=1}^{K} m_k(t)e^{j2\pi f_0 \tau_i(\theta_k + \phi_k)} + n_l(t). \quad (9)$$

Time required signal which up to elemen $i$ from receiver array from reference array element with signal $k^{th}$ which arrive from $(\theta_k, \varphi_k)$ can be calculated as [10]

$$\tau_l(\theta_k, \phi_k) = \frac{r_l \cdot v(\theta_k, \phi_k)}{c} \quad (10)$$

where $r_l$ is the position vector of antenna $i$ and $v(\theta_k, \varphi_k)$ is the unit vector which aim to signal of arrival $k$. Autocorrelation matrix from array sensor can be obtained from [10]

$$R = E\{XX^T\}. \quad (11)$$

The correlation matrix of received signal from antenna element can be found from [10]

$$R = \frac{1}{K} \sum_{n=1}^{N} x_n x_n^H. \quad (12)$$

To calculate eigenvalues and eigenvectors from correlation matrix using (12), input a noise subspace matrix where eigenvectors which suited for smallest eigenvalues from correlation matrix.

For every $\theta$ and $\varphi$, create steering vector with using equation that given by [10]

$$s(\theta, \phi) = \begin{bmatrix} e^{j2\pi f_0 \tau_1(\theta, \phi)} & \cdots & e^{j2\pi f_0 \tau_l(\theta, \phi)} \end{bmatrix}. \quad (13)$$

Calculating $P_{MU}$ for each angle using equation to find peak from MUSIC spectrum which estimated angle DOA [7]

$$P_{MU}(\theta, \phi) = \frac{1}{\left| s^H(\theta, \phi).U_L \right|^2} \quad (14)$$

where $U_L$ is a matrix with the dimension $l \times l$–$m$ as eigenvectors which appropriate with smallest of $l$–$m$ eigenvalues from correlation matrix of array [7].

III. SIMULATION RESULTS

System for estimation signal in inter-satellite-link system performed with doing constellation satellite simulation where $K$ satellite want to communicate with one main satellite. Each of satellite will have information of azimuth and elevation position. But in this case only azimuth position that will be taken and the main satellite only detect for 1800 view. Later each satellite will emit an electromagnetics wave with power that has been decided.

Evaluation scenario for MUSIC algorithm in inter-satellite-link system will be reviewed from several aspects first. First, how to calculate and find the values of maximum angular beamwidth for MUSIC algorithm in inter-satellite-link case. Second, how the number of AOA changes toward output graphic from MUSIC algorithm. Third, how the number of element changes toward output graphic. Fourth, how the number of frequency operation changes toward output graphic. Fifth, how spacing between elements changes toward graphic output. Sixth, how different power receive changes toward graphic output. Last, how to calculate computing time for MUSIC algorithm. Later, computing time for MUSIC algorithm will be compared to time of satellite when moves in orbit.

*A. Maximum Angular Beamwidth*

Calculating maximum angular beamwidth addressed to find out the threshold detection of MUSIC algorithm. So, if applying maximum angular beamwidth to ISL system, calculating maximum angular beamwidth can determine maximum number of satellite constelation so the main satellite can receive and detect signal of arrival from another satellite accurately.

Fig. 2 shows values of maximum angular beamwidth can be decided in MUSIC algorithm. In Fig. 2 (a) obtained simulation result graph for maximum angular beamwidth from 0 until 180 degree. The graph is using step width of 1 degree and it still can detect the angle of arrival accurately with differences of one degree stepwidth. In Fig. 2 (b) to see more clearly average peak power MUSIC for every AOA which detected by MUSIC algorithm with range angle from 60 until 80 degree, it detects 20 angle of arrival (AOA).

*B. Increasing Number of Angle of Arrival*

For this analysis will performed graph of comparison result from one AOA until 20 AOA. The comparison that will be performed is values of PMUSIC(dB) and accuracy from graph.

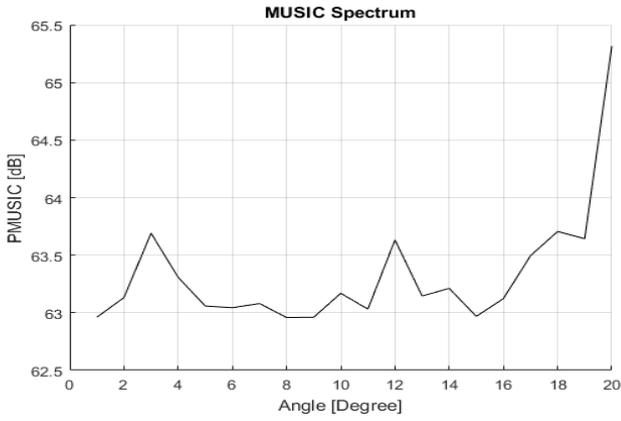

Fig. 3. Simulation Results of Increasing AOA to PMUSIC(dB)

Fig. 3 shows how increasing AOA affect to MUSIC algorithm. For increasing AOA occur fluctuations. The graph shows that when MUSIC algorithm detects one AOA, MUSIC algorithm receives a signal around 62.9 dB.

When the number of AOA increased become two, it happens increasing receive a signal around 63.1 dB. When the number of AOA increased become three, it happens increasing receive a signal around 63.6 dB, and so with others number of AOA, happened fluctuations for every increased the number of AOA. So, not every increased number of AOA will decrease PMUSIC that received by MUSIC algorithm.

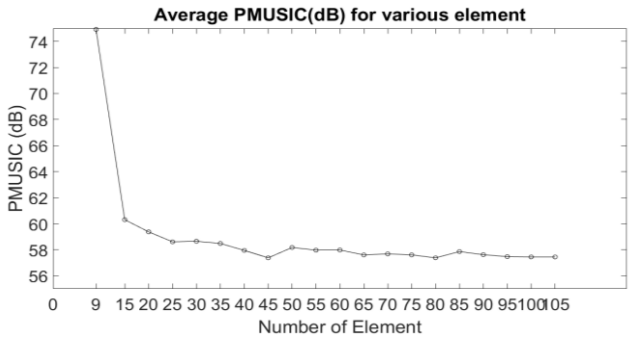

(a)

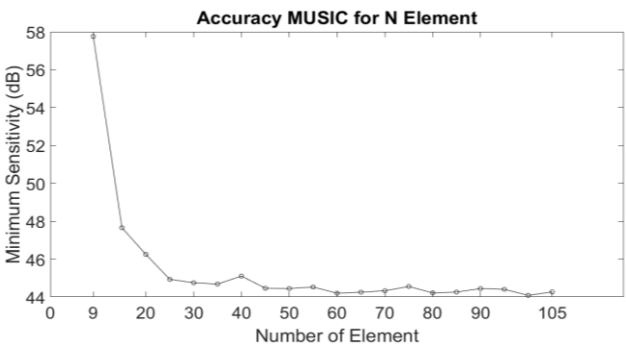

(b)

Fig. 4. Simulation Results of Comparison Number of Element Antenna: (a) for Average Peak Power, (b) for Minimum Sensitivity

### C. Increasing Number of Antenna Element to Average Peak Power Music and Accuration Level

In this analysis will be performed graph of comparison from nine to 105 elements of antenna. Fig. 4 (a) depicts that increasing element of antenna will decrease PMUSIC. But for better and more accurate decision, the effect of increasing number of antenna element should be reviewed in minimum sensitivity detection. Fig. 4 (b) depicts that the minimum sensitivity is increased along increasing number of antenna element. Minimum sensitivity can be a parameter to detect accuracy level, because for certain PMUSIC, angle of arrival has been detected. For 9 elements of antenna, MUSIC algorithm can detect AOA with PMUSIC level around 60 dB. For 20 elements of antenna, MUSIC algorithm can detect AOA with PMUSIC level around 46 dB. When the number of antenna elements is increased, it will produce lower PMUSIC level to detect AOA accurately. It can be concluded that more accurate to detect AOA from a signal. It produces lower PMUSIC level which be used by MUSIC algorithm.

### D. Different Spacing between Element Antenna

For this analysis will be performed for different spacing from $0.25\lambda$ until $5\lambda$ with number of antenna elements is 50 and the operating frequency is 32 GHz which produce the wavelength of 9.4 mm. In Fig. 5, given information that detection accuracy level of AOA for spacing 5 mm or $0.5\lambda$ with accuracy level of 95 percent. For increased spacing is occur fluctuation of graph, but for every increased AOA for accuracy level always below 50 percent. The level of accuracy increased again when the spacing equals to 40 mm with 70 percent of accuracy, and decreased when the spacing equals to 47.5 mm. This phenomenon can be happened because the greater spacing between antenna in MUSIC algorithm, the greater MUSIC algorithm will detect subspace source so there will be unwanted signal which detected but not greater as AOA signal.

### E. Different Operating Frequency

This analysis will perform effect of different frequency operation for MUSIC algorithm. Frequency operation that will be compared is 23 GHz, 24.5 GHz, 32 GHz, and 56 GHz corresponding to the Table 1. [8]

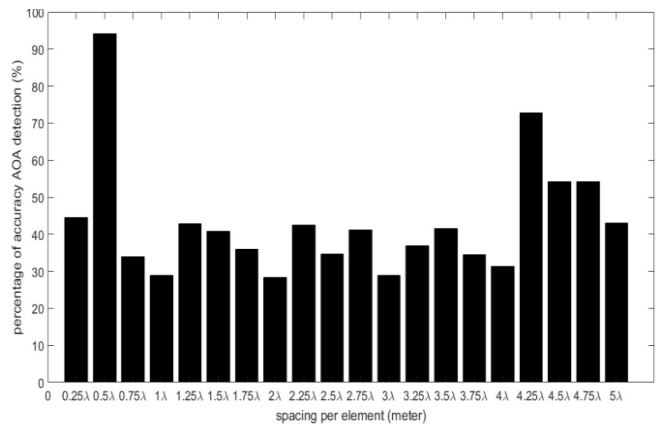

Fig. 5. Simulation Result for Different Spacing

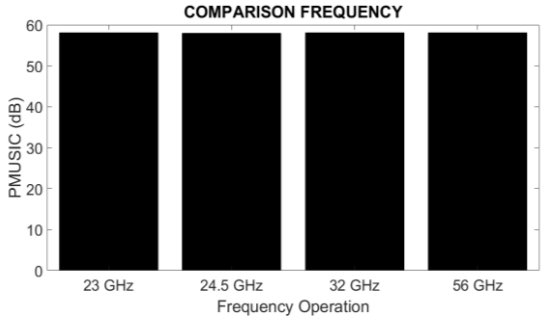

Fig. 6. Simulation Result for Different Frequency Operation

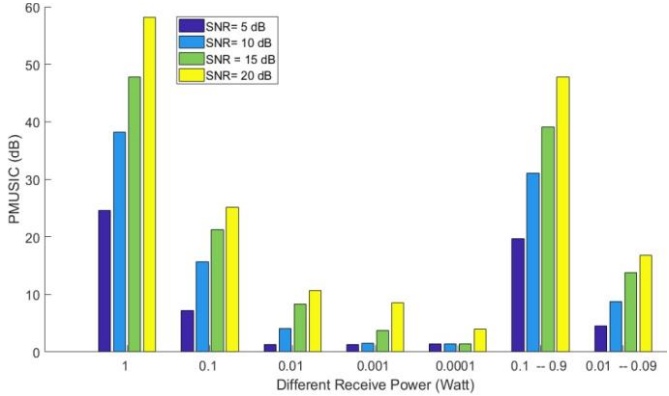

Fig. 7. Simulation Result for Different Receive Power in Every Antenna Element at SNR = 5 dB, SNR = 10 dB, SNR = 15 dB, SNR = 20 dB

TABLE I.  TABLE STYLES

| Intersatellite service | Frequency Bands |
|---|---|
| Radio Frequency | 22.55 GHz – 23.55 GHz |
|  | 24.45 – 24.75 GHz |
|  | 32 – 33 GHz |
|  | 54.25 – 58.2 GHz |

In Fig. 6, given information that the largest PMUSIC(dB) is when the frequency operation is 32 GHz. But the changes of PMUSIC(dB) value for every frequency operation is not too significant, because for each frequency operation produce PMUSIC(dB) with range around 57 dB and 58 dB.

*F. Different Receive Power in Each Antenna Element*

This analysis performs the effect of difference receive power in every antenna element for each AOA. For noise signal amplitude amount of 50 watt. Number 1 indicate power receive that received by MUSIC algorithm amount of 1 watt, number 2 amount of 0.1 watt, number 3 amount of 0.01 watt, number 4 amount of 0.001 watt, number 5 amount of 0.0001 watt, number 6 amount of range power from 0.1 watt to 0.9 watt, number 7 amount of range power from 0.01 watt to 0.09 watt. In Fig. 7 (a), given information that power receive amount of 1 watt or 0 dB.

MUSIC algorithm can detect AOA with PMUSIC(dB) around 24 dB. When power receive is decreased to 0.1 watt or -10 dB, the PMUSIC(dB) decreased around 6 dB. Although there is decreasing PMUSIC(dB), it still can detect AOA accurately. but, if PMUSIC(dB) below 2 dB, MUSIC algorithm will not detect AOA accurately, so for 0.01 watt or -20 dB, 0.001 watt or -30 dB, 0.0001 watt or -40 dB, and 0.01 – 0.09 watt or -10.4576 – 20 dB. The result can be happened because in MUSIC algorithm, every signal of arrival will be divided into 2 parts, the first one is subspace source and the second one is subspace noise. Where subspace source and subspace noise are data matrix which consist of complex number. For taking real number from each subspace, the value can be compared with adding every real number of the matrix. If total real number from subspace source is greater than total real number from subspace noise, the MUSIC algorithm will detect AOA accurately.

The solution to increase accuracy level in 0.01 watt, 0.001 watt, 0.0001 watt, and 0.01–0.09 watt is to increase SNR. Increasing SNR will minimize noise variance in AWGN channel as can be seen in Fig. 7 (b).

*G. Comparison for Computing Time to Satellite Velocity*

For computational time calculation analysis, the first calculation is done when the SPOT satellite are used for earth imagery observation. The satellite moves with an altitude of 830 km in a period of 101 minutes. However, the lack of information about SPOT satellites is the apogee and perigee location so it is difficult to calculate the circumference of SPOT satellites in elliptical form. In this case it is assumed for circular orbits of satellites. The first scenario is to calculate the circumference of the circular orbit of the SPOT satellite.

Calculation of the circumference of SPOT satellite orbit requires the radius of the earth and the height of the SPOT satellite, where the SPOT satellite altitude is 830 km and the radius of the earth is 6,371 km, then the radius of the circumference is 7,201 km and the circumference of the SPOT satellite orbit is 45,222 km. Furthermore, the SPOT satellite speed is calculated with the formula $v = s/t$, where $s$ is the circumference of SPOT satellite orbit and $t$ is SPOT satellite period of 101 minutes or 1.683 hours. The obtained SPOT satellite speed is 26,870 km/h or 7,463.9 m/s. The next step is calculation the distance per one degree of SPOT satellite orbit, where the circumference of SPOT satellite orbit is 360 degrees, then the obtained distance per one degree is 125.617 km. Thus, the time required by SPOT satellite to travel a path of one degree in orbit is 16.83 seconds.

Fig. 8 describes that increasing number of AOA and number of antenna element in all of operating frequency will increase computation time. On the contrary, the less received power, the less computation time. Meanwhile increasing spacing between antenna elements give fluctuation but not too significant to affect the computation time. Comparison also were made whether the largest computational time of the MUSIC algorithm was slower than orbit interval in one degree. When the MUSIC algorithm's largest computing time is below the orbit interval in one degree, the MUSIC algorithm matches to the Inter-Satellite Link requirement. The largest computational time is about 0.98 seconds, where the orbit interval is 16.83 seconds, so the MUSIC algorithm is suitable for Inter-Satellite-Link system.

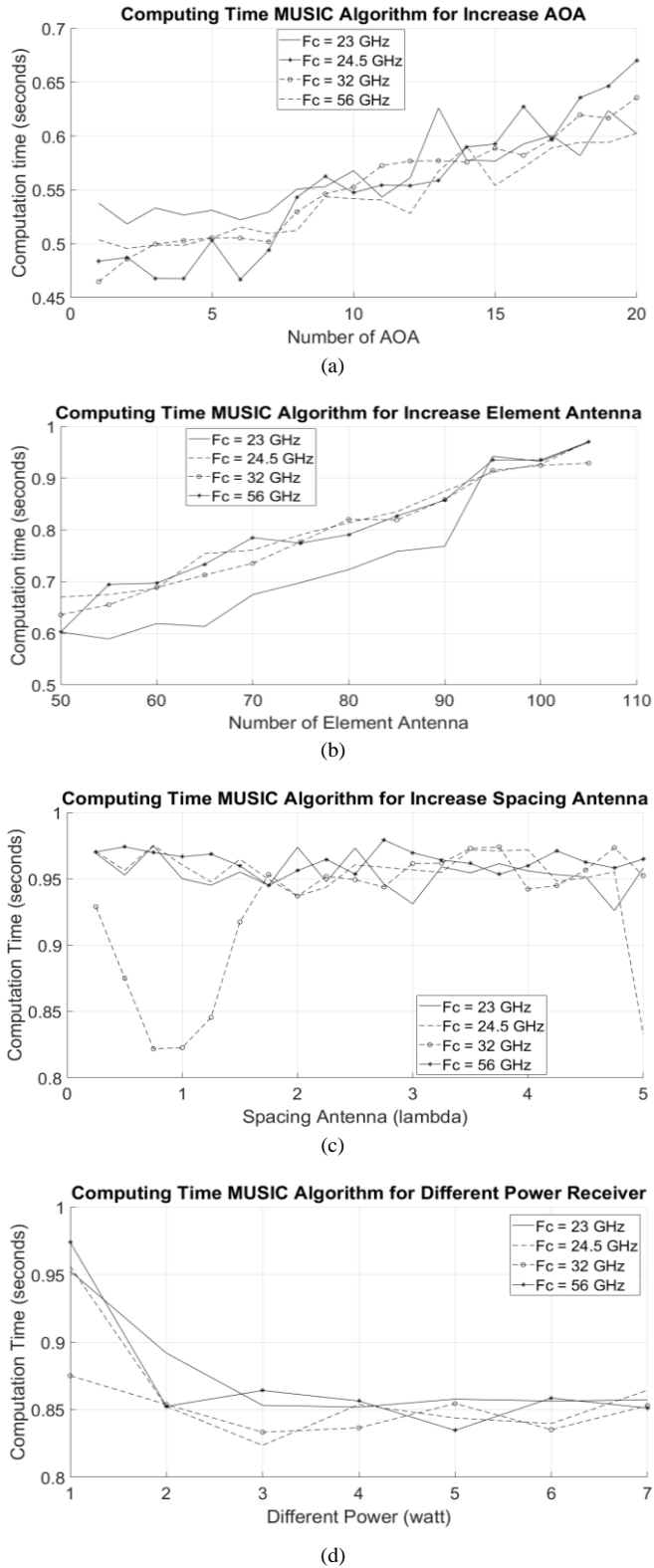

Fig. 8. Simulation Eesult for Computation Time: (a) Increasing AOA . (b) Increasing Number of Element Antenna. (c) Increasing Spacing Antenna. (d) Different Power Receive

IV. CONCLUSION

The detection of Direction of Arrival (DoA) using the MUSIC algorithm can accurately detect the direction of signal arrival or angle of arrival (AoA) based on the analysis of simulation results on several parameters. The largest improvement effect by the MUSIC algorithm in DoA detection is the increasing in the number of antenna elements, followed by increasing received power level which increase the SNR, increasing the number of AOA detection, increasing inter-elements spacing, and increasing frequency. The conclusions are valid as long as appropriate with the limitation of volume and mass budget associated to the total dimension and weight of the array antenna. The use of 50 elements array antenna with 0.5 wavelength spacing between adjacent elements, the MUSIC algorithm can correctly detect 20 angles of arrival (AoAs). The total dimensions of array antenna are 65.22, 61.22, 46.88, and 26.79 cm in each of the operating frequencies 23, 24.5, 32, and 56 GHz.